# Chitosan Nanocomposites with CdSe/ZnS Quantum Dots and Porphyrin

F. A. Sewid [a,b,*], K. I. Annas [a], A. Dubavik [a], A. V. Veniaminov [a], V.G. Maslov [a], A. O. Orlova [a*]

[a] ITMO University, St. Petersburg197101, Russia

[b] Faculty of Science, Mansoura University, Egypt

*Corresponding Authors

*e-mail: fayzaomar8891@gmail.com

*e-mail: a.o.orlova@gmail.com

**Abstract**

Water-soluble nanocomposites based on CdSe/ZnS quantum dots (QDs) and hydrophobic tetraphenylporphyrin (TPP) molecules passivated by chitosan (CS) have been formed. Magnetic circular dichroism (MCD) spectra evidence TPP presence in both monomeric and agglomerated forms in the nanocomposites. The nanocomposites demonstrate more pronounced singlet oxygen generation in comparison with free TPP in CS at the same concentration due to intracomplex Förster resonance energy transfer (FRET) with 45 % average efficiency.

**Introduction**

In the last decades, numerous studies have been devoted to finding an effective approach to cancer treatment. Photodynamic therapy (PDT) out performs the conventional cancer treatment methods, being a non-invasive, inexpensive therapy approach, with low toxicity, no drug resistance, and fewer side effects[1–3]. PDT employs organic photosensitizers (PSs) capable of producing singlet oxygen (SO) by energy transfer to triplet oxygen molecules upon exposure to light. In turn, SO is responsible for killing cancer cells [4–7].

Photosensitizers (PSs), especially porphyrin, and their derivatives, are mostly used for PDT due to their unique properties. Some of these photosensitizers have been approved for clinical practice. The drawbacks of these photosensitizers include poor solubility in water, aggregation in aqueous solutions, narrow absorption bands, the release of phototoxic products, and slow elimination from the body[8,9].

Nanomaterials can overcome the limitations of organic photosensitizers due to their stability under irradiation and enhanced PSs functional properties[10]. Quantum dots (QDs) are extensively studied for PDT applications among different nanomaterials. QDs possess unique properties, including high photoluminescence quantum yield (PLQY), long PL lifetimes, high extinction coefficients, size-dependent spectral properties, superior chemical stability, photostability, and high surface activity, as well as facile surface modification[11–14]. Subsequently, QDs conjugated with porphyrin allow the formation of QD-porphyrin hybrid structures with superior photophysical properties, including an increase in singlet oxygen generation[15–17]. QDs are also used as donors of Forster resonance energy transfer (FRET) in these structures. The application of a QDs-porphyrin hybrid structure is usually greatly hampered by its toxicity [18]. A combination of porphyrin with QDs, incorporated in cationic biopolymers, yielded the water-soluble and biocompatible nanocomposite with improved delivery and distribution in the cells[19,20].

A new trend of chitosan (CS) biopolymer application as a stabilizing agent and a carrier for photocatalysts, photosensitizers, and quantum dots has been developed in recent years[21,22]. CS is a natural polysaccharide derived from natural sources; it has unique properties, including biocompatibility, hydrophilicity, biodegradability, nontoxicity, and chemical stability[23,24]. These properties make CS appropriate for many biomedical applications, including drug delivery systems, tissue regeneration, and artificial skin[25,26]. CS is used as a super-adsorbent for porphyrin molecules[27] and strongly attracts negatively charged nanoparticles, e.g., quantum dots stabilized by thioglycolic acid (TGA)[28]. The composition of QDs and porphyrin molecules with biodegradable polymeric carriers makes them biocompatible for medicine. CS reduces nanocomposites aggregation in a biological environment and increases interactions with biological molecules[29]. Nanocomposites of QDs with porphyrin will show promising results in PDT in transferring the hybrid structures to aqueous media while retaining their photophysical properties.

This paper discusses the formation of water-soluble nanocomposites based on hydrophilic CdSe/ZnS QDs and hydrophobic TPP in CS solution. The incorporation of QDs-TPP nanocomposites in CS has significant attractive features, such as making these structures biocompatible and keeping their photophysical properties. QDs-TPP nanocomposites are distinguished by more efficient SO generation compared to free TPP in CS due to efficient FRET from QDs to TPP.

## EXPERIMENTAL SECTION

### Materials

All chemical reagents were analytical grade and used as purchased without further purification. TPP was purchased from Frontier Scientific (USA). Chitosan with low Mw (50-190 kDa) and the deacetylation degree (DD) 75%, acetic acid, potassium hydroxide (KOH), and tetrachloromethane ($CCl_4$) were purchased from Sigma-Aldrich (USA). The chemical sensor (singlet oxygen sensor green, SOSG) was purchased from Frontier Scientific (USA).

### Ligand Exchange of CdSe/ZnS QDs

CdSe/ZnS core/shell QDs with an average core size of 3.4 nm were stabilized with oleylamine and trioctylphosphine (TOP). The synthesis of CdSe/ZnS QDs was performed according to the protocol described in[30]. A solution of as-prepared QDs (~5 mg) in chloroform (250 μL) was washed with methanol (500 μL) to remove an excess of oleylamine and trioctylphosphine (TOP), then the sample was centrifuged (1 min, 15000 rpm), and dissolved in chloroform (750 μL). Initial hydrophobic ligands were replaced with hydrophilic ligands, i.e. thioglycolic acid (TGA), by adding a concentrated solution of sodium thioglycolate (75 μL) in methanol (10 mg/ μL) to the QDs solution in chloroform. The mixture was intensively mixed and left for 2 minutes, and after that, QDs were stabilized with TGA. 85 mg of KOH was dissolved in 1.5 mL of water, and 790 μL of the solution was added to the mixture. The resulting mixture was carefully stirred (slowly turning the mixture upside down and back again). As the result, QDs were transferred to the water.

### Preparation of QDs-TPP nanocomposites

The nanocomposites were formed by dissolving free CS in aqueous acetic acid (0.1% w/w in 0.1% aqueous acetic acid) under magnetic stirring for ~24 h until obtaining a CS solution. pH of the CS solution was adjusted at ~ 5.5-6, and then QDs stock solution ($C_{QDs}$ ~ $3 \cdot 10^{-7}$ M) was added; the mixture was left on the stirrer for 2 hours. After that, TPP solutions in $CCl_4$ ($C_{TPP}$ ~ $(1.7 \div 11) \cdot 10^{-7}$ M) were added to the mixture and stirred to evaporate $CCl_4$ and achieve the sorption–desorption equilibrium. Finally, the nanocomposites were formed (see the proposed model is presented in Figure S1 in SI file 2) with molar ratio $n = C_{TPP}:C_{QDs}$ ranging from 0 to 4. CS is a polycationic biopolymer with two functional groups, hydroxyl and amine groups (OH and $NH_2$), respectively. The most probable formation of nanocomposites resulted from the electrostatic interaction between CS functional groups and the stabilizer molecules TGA on the

QDs surface[31]. We suppose that QDs-TPP nanocomposites are formed because of the coordination of TPP on the QDs surface[15].

*Estimation of the efficiency of singlet oxygen generation by TPP and QDs-TPP nanocomposite in chitosan solution*

Singlet oxygen generation efficiency was measured using the Singlet Oxygen Sensor Green chemical sensor (SOSG)[32]. The samples were mixed with SOSG 5 mM solution in water with methanol and irradiated by 460 nm LED with uniform 3 mW/cm² power density. At each specific time point, the PL of SOSG was detected at 530 nm when it was excited at 500 nm. To study the generation of SO, SOSG PL was monitored by measuring its intensity after each specific portion of external radiation. TPP was used as a sensitizer of SO generation, in a free state and with QDs in the nanocomposite in CS.

Singlet oxygen generation quantum yield $\phi_\Delta$ by nanocomposite was estimated using a comparative method under ambient conditions, and the calculations were done according to the literature [33,34]. Ce6 was used as a standard photosensitizer reference with $\phi_\Delta = 0.64$ in water at pH ~ 8[35].

*Optical characterization of CdSe/ZnS QDs -TPP nanocomposites*

Optical properties of nanocomposites and their components were studied using a Shimadzu UV-3600 spectrophotometer and Cary Eclipse (Varian) fluorescence spectrophotometer, respectively. All measurements were carried out at room temperature. A quartz cuvette with1 cm path length was used for all spectroscopic measurements. Magnetic circular dichroism (MCD) spectra were recorded by Jasco J-1500 CD spectrophotometer and MCD-581 electromagnet with a 1.5 T field strength. The average particle size and Zeta potential were detected using the Malvern Zetasizer Nano model ZS90 system. PL kinetics of the samples has been analyzed using the time-resolved fluorescence microscope MicroTime100 (Pico Quant, Berlin, Germany). The Zeiss continuous interference filter monochromator with 10 nm FWHM was used to select the PL intensity of QDs in the nanocomposite samples. The PL decay curves of QDs and the nanocomposite were fitted with a biexponential function:

$$y = y_0 + A_1 \cdot \exp\left[-\frac{(X - X_0)}{t_1}\right] + A_2 \cdot \exp\left[-\frac{(X - X_0)}{t_2}\right] \qquad (1)$$

where $A_i$ and $t_i$ are the amplitude and decay time of the $i$ th (1st or 2nd) component.

The average PL decay time, $\langle \tau \rangle$, was calculated according to the Equation:

$$\langle \tau \rangle = \frac{A_1 \tau_1^2 + A_2 \tau_2^2}{A_1 \tau_1 + A_2 \tau_2} \tag{2}$$

## *Results and Discussion*

Our previous study has demonstrated the formation of QDs-TPP hybrid structures in organic medium (CCl$_4$) with different molar ratios with high SO generation efficiency due to QD-TPP FRET. There are no aggregation of TPP in structures with QDs were found there[15]. In this paper, we present results of study of the photophysical properties of water-soluble QDs-TPP nanocomposites in CS.

High FRET efficiency can be achieved in QDs-TPP nanocomposites because of the spectral superposition between the QD emission and Q absorption band of TPP molecules in CS solution shown in Figure 1.

Figure 2 shows the absorption and PL spectra of CdSe/ZnS QDs dispersed in CCl$_4$, water, and CS solution. The first excitonic absorption band of QDs is detected at 564 nm, and the average core diameter of CdSe QDs is estimated as 3.4 nm, according to the literature [36]. As shown in Figure 2, the position of PL band of QDs does not change due to transfer QDs from CCl$_4$ to water. It confirms that QDs are stable in different media.

Figure 3 exhibits absorption spectra of QDs, TPP, and QDs-TPP nanocomposites in CS solution with different molar ratio (*n*). Comparing the TPP absorption spectrum in CCl$_4$ (see Figure S1 in SI file 1) with its spectra in CS and nanocomposites, it is noticeable that the positions of TPP bands in the nanocomposites and CCl$_4$ are the same. At the same time, broadening of the Soret band of TPP absorption spectrum in CS is a pattern of the formation of non-luminescent H-type aggregates of TPP[37].

Dynamic light scattering (DLS) measurements display that the hydrodynamic size of QDs is 20 nm, and the QDs-TPP nanocomposite is 24.5 nm (see Figure S2 in SI file 1). The zeta potential of free CS at pH 5.5, QDs after incorporation within CS and QDs-TPP-CS nanocomposite are +42 mV, +44 mV, and +39.8 mV, respectively, and zeta potential of free

QDs in water is -25 mV (see Figure S3 in SI file 1). The zeta potential of the nanocomposite decreased due to the aggregation of TPP in CS solution.

It should be pointed out that the absorption spectra of the samples did not provide any evidence for energy transfer or interaction between QDs and TPP. MCD spectroscopy has proved to be a powerful tool to probe the samples' electronic structures [38,39]. Therefore, MCD spectroscopy has been applied to study TPP interaction with QDs and CS in an aqueous solution and detect the non-luminescent TPP aggregates in samples.

MCD spectra of the polysaccharide in an aqueous solution at pH 5.5 have a negative band at 190-250 nm spectral range (see Figure S4 in SI file 1), related to n → π* transition [40,41]. Therefore, MCD spectra of CS did not overlap with the bands of QDs or TPP due to CS transition in the far-UV. MCD spectra of CdSe QDs in CS presented in Figure 4a have shown a series of bands, including A and B terms, and reflected the diamagnetic properties of CdSe QDs[42].

MCD spectra of free TPP monomers in $CCl_4$ (Figure 4b) comprise two opposite-signed symmetrical bands centered at 416 nm and 425 nm due to Zeeman splitting of the Soret band. The Soret band should be very sensitive to any interaction between TPP molecules and their interactions with an aqueous medium and QDs[43,44]. The presence of TPP aggregates in CS is confirmed in Figure 4c by two opposite-signed spectra with additional small shoulders and low symmetrical signals at the Soret band between 400 to 435 nm; the center of the Soret band is at 423 nm, as is shown in absorption spectra in the bottom panel due to Zeeman splitting and B term of the Soret band and Q bands. A shoulder at 399 nm might be related to TPP aggregates in CS solution[45]. MCD spectra of the TPP Soret band with QDs in the nanocomposite shown in Figure 4d have the same bands as for TPP in CS solution. It confirms the presence of TPP monomers and aggregates in the nanocomposite.

The PL decay curves of free CdSe/ZnS QDs, free TPP, and QDs-TPP nanocomposite in CS solution are presented in Figure 5. The notable decrease in the PL intensity of QDs in nanocomposite compared to free QDs demonstrates the formation of QDs-TPP nanocomposites due to FRET from QDs to TPP monomers and their aggregates. The PL lifetimes and amplitudes of the QD PL components have shown that QDs in both PL fractions are quenched in nanocomposite due to efficient energy transfer from QDs to TPP molecules [46], as shown in Table

1. The shortening of TPP PL decay time in CS (see Figure S3 in SI file 2) may indicate the increase in the nonradiative S1→S0 transition rate in comparison with TPP monomers in $CCl_4$.

Figure 6 shows PL spectra of CdSe/ZnS QDs and QDs-TPP nanocomposites with different molar ratio excited at 460 nm. Only QDs could be excited at 460 nm because of the negligible absorbance of TPP at this wavelength (See Figure 3). The appearance of PL of TPP in mixed solutions indicates the efficient FRET from the excited QDs to TPP molecules. Thus, it suggests the formation of QDs-TPP nanocomposites.

Also, we have analyzed the PL excitation spectra (PLE) of the samples shown in Figure 7 to estimate the FRET efficiency, confirming the formation of the QDs-TPP nanocomposites in CS. The FRET model[47] was utilized to estimate the energy transfer efficiency in nanocomposites. The contribution of the absorption spectrum of QD to the TPP PLE spectra (see the region covered with the rectangle) is the direct evidence for the FRET from QDs to TPP molecules. The efficiency of FRET from QDs to TPP is approximately 45% at the lowest TPP concentration in the samples. An increase in the TPP concentration in nanocomposites led to a sharp decrease in FRET efficiency from 45 % to 17 %. It indicates that the increase in the concentration of TPP leads to an increase in non-luminescent TPP aggregates in the mixture with QDs. Non-luminescent TPP aggregates can be efficient acceptors of energy and they can quench both phosphors in our system, i.e. QDs and TPP monomers [48].

The efficiency of the SO generation by QDs-TPP nanocomposite and free TPP with the same TPP concentration in CS solution has been analyzed to study if the nanocomposites can generate the SO after their incorporation in CS biopolymer.

*Singlet oxygen generation by QDs-TPP nanocomposite and free TPP in chitosan solution*

To monitor SO generation, the samples of CS with free TPP and QDs-TPP nanocomposites with a molar ratio of 0.5 and with the same TPP concentration were irradiated with visible light (460 nm) at room temperature, and SOSG PL spectra were measured after each portion of irradiation (see Figure 8. a and 8. b). Judging from the changes in SOSG PL intensity at 530 nm are shown in Figure 8c, the SO concentration reaches the maximum value at the irradiation doses of 6.12 and 5.04 J in the nanocomposite and free TPP, respectively, and then remains at a saturation state. Moreover, a comparison of the SOSG PL intensity in nanocomposite with free TPP samples, shows that TPP monomers in nanocomposite with QDs generate SO more efficient than TPP in CS due to FRET from QDs to TPP monomers. It has been found that TPP monomer's concentration in our nanocomposite at the molar ratio ($n= 0.5$) is equal to 40% of the total TPP concentration (see formula 2 in SI file).

Concentration of SO generated by TPP monomers in CS under the external radiation can be determined as:

$$C_{SO}^{direct} \sim \varepsilon_{TPP} C_{TPP} E_{SO}^{TPP} \tag{3}$$

where $\varepsilon_{TPP}$ and $C_{TPP}$ are the extinction coefficient and the TPP monomer concentration; $E_{SO}^{TPP}$ is the efficiency of SO generation by TPP molecules.

Concentration of SO generated by TPP monomers in the nanocomposite due to FRET from QDs to TPP can be determined as:

$$C_{SO}^{sens} \sim \varepsilon_{QDs} C_{QDs} E_{FRET} E_{SO}^{TPP} \tag{4}$$

where $\varepsilon_{QDs}$ and $C_{QDs}$ are the extinction coefficient and the concentration of QDs and $E_{FRET}$ is the FRET efficiency from QDs to TPP.

Concentration of SO generated in the samples is proportional to the change in SOSG PL intensity at 530 nm when SO interacts with the SOSG sensor.

$$C_{SO}^{direct} \sim \Delta I_{SOSG}^{direct}; C_{SO}^{sens} \sim \Delta I_{SOSG}^{sens} \tag{5}$$

where $\Delta I_{SOSG}^{direct}$ and $\Delta I_{SOSG}^{sens}$ are the change in SOSG PL intensity for direct and sensitized generation, respectively.

The dependences presented in Figure 8 clearly demonstrate that QDs-TPP nanocomposite generates SO in larger amounts than free TPP because of FRET from QDs to TPP molecules. We believe that SO quantum yield of TPP monomers in nanocomposites and CS is the same. Then, we have used the PL intensity of SOSG to estimate FRET efficiency in our samples as follows from Eqs. (3)-(5), where $f$ is the fraction of QDs associated with TPP monomers. The FRET efficiency can be calculated as:

$$E_{FRET} = \frac{\Delta I_{SO}^{sens} \cdot C_{TPP} \cdot \varepsilon_{TPP}}{\Delta I_{SO}^{direct} \cdot C_{QDs} \cdot \varepsilon_{QDs} \cdot f} \tag{6}$$

According to Eq. 6, the average FRET efficiency in the QDs-TPP nanocomposite is $25\pm2\%$.

Comparison of FRET efficiency in our composites estimated from QD quenching and SO generation demonstrates that only 50% of energy transferred from QD to TPP uses for SO production. It confirms FRET from QDs to non-luminescent TPP aggregates. Nevertheless, our data clearly demonstrate that combination of QDs and TPP in nanocomposites in CS leads to 1.5 times rises of SO production in comparison with free TPP.

*Conclusion*

In summary, the nanocomposite based on CdSe/ZnS QDs and hydrophobic TPP molecules in CS has been formed. We successfully transfer to water and save approximately 40% of its in monomeric form. As a result, TPP monomers in CS are able to generate SO with the same QY as TPP monomers in $CCl_4$. We demonstrate that introduction QDs in CS with TPP allows to twice increase SO production compared to free TPP in CS. The nanocomposites are characterized by excellent stability and efficient SO production due to efficient QD-TPP monomer FRET. Consequently, we believe that these nanocomposites might be perfect SO generators in PDT.

*Acknowledgements*

The research was supported by the Ministry of Education and Science of the Russian Federation, State assignment, Passport 2019-1080 (Goszadanie 2019-1080).The researcher F. A. Sewid is partially funded by a scholarship under the joint (Executive Program between the Arab Republic of Egypt and Russia)

Table 1. Lifetime data analysis of QDs' PL in CS and nanocomposite.

| Samples | $\tau_1$, ns | $\tau_2$, ns | $A_1$ | $A_2$ | $\langle\tau\rangle$, ns |
|---|---|---|---|---|---|
| QDs | 16±0.3 | 4.1±0.21 | 946±22 | 1840±92 | 8±0.2 |
| TPP | 5.8±0.11 | ----- | 170±5.2 | ---- | 5.8±0.11 |
| QDs in nanocomposite | 14±0.7 | 3.6±0.18 | 211±12 | 526±26 | 6.5±0.1 |

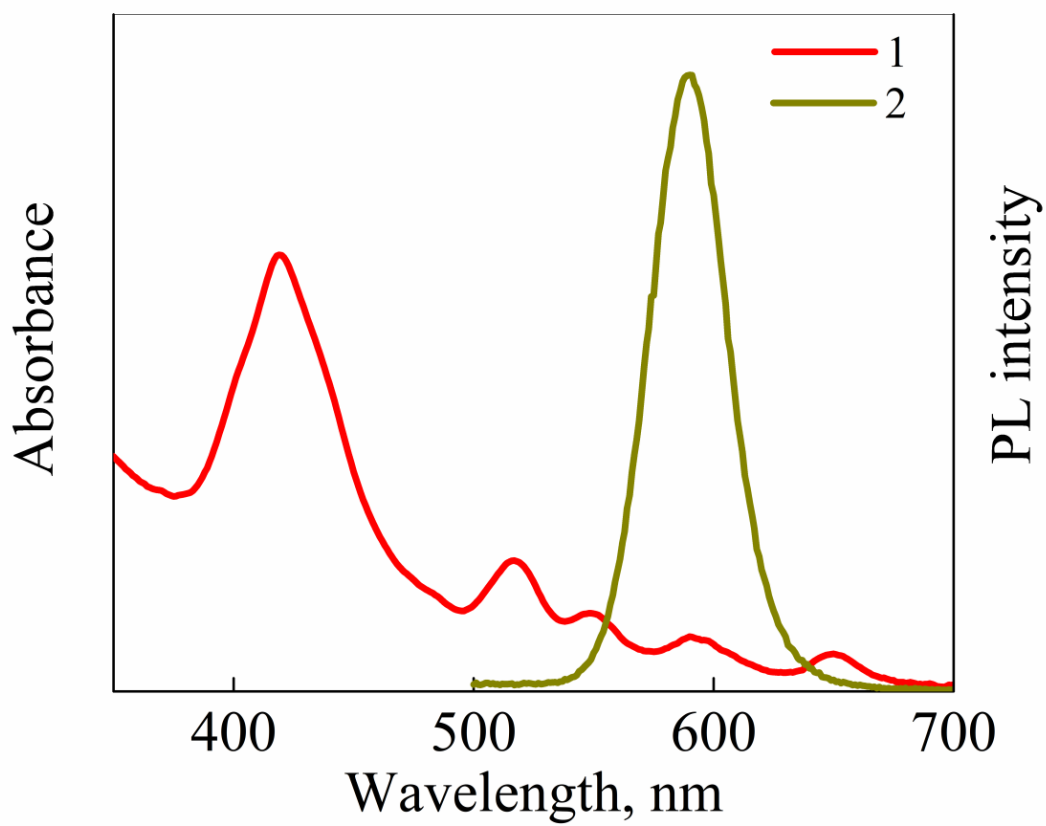

Figure 1. (1) Absorption spectrum of TPP and (2) PL spectrum of CdSe/ZnS QDs in CS solution. PL excitation wavelength is 460 nm.

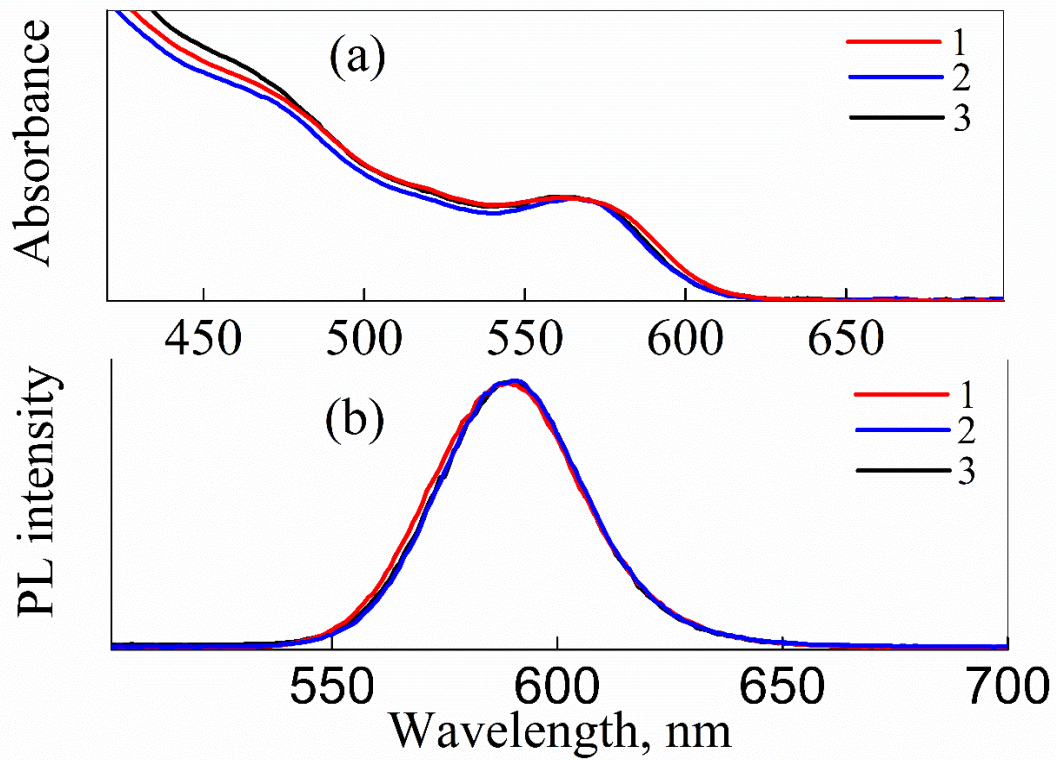

Figure 2. Absorption (a) and PL (b) spectra of CdSe/ZnS QDs dissolved in $CCl_4$(1), water (2), and CS (3); PL excitation wavelength is 460 nm.

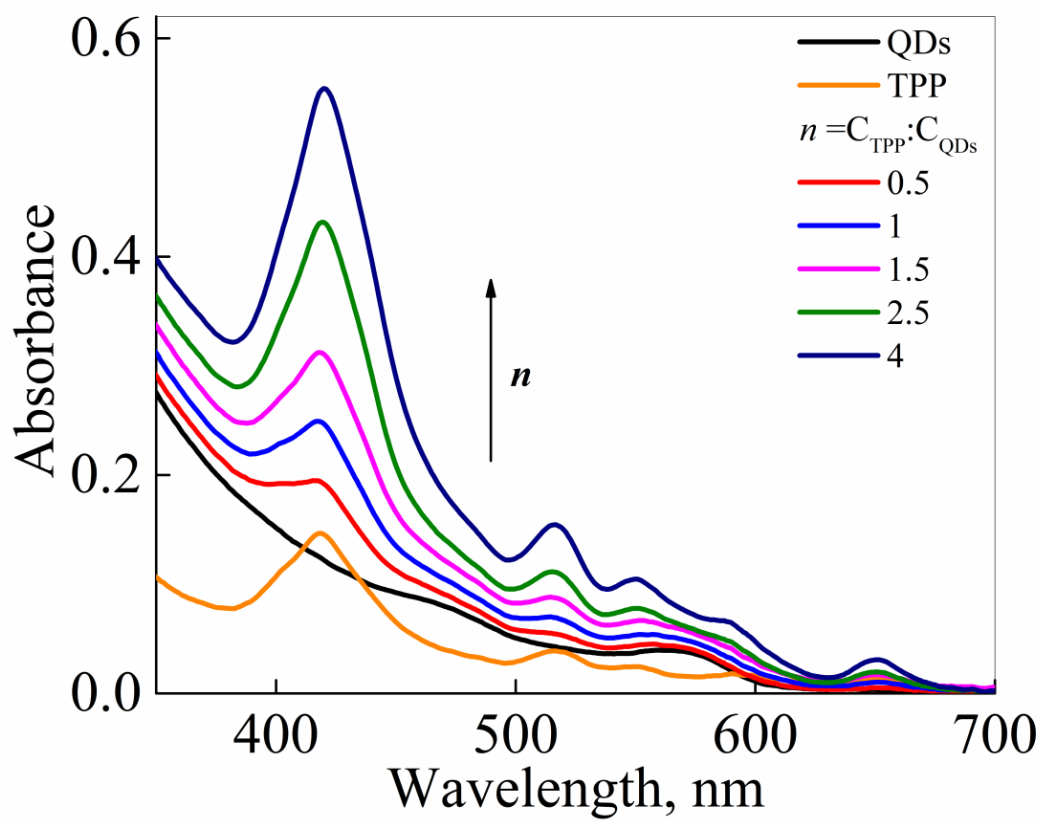

Figure 3. Absorption spectra of QDs, TPP, and QD-TPP nanocomposites in CS with different molar ratio *n* values (0.5- 4).

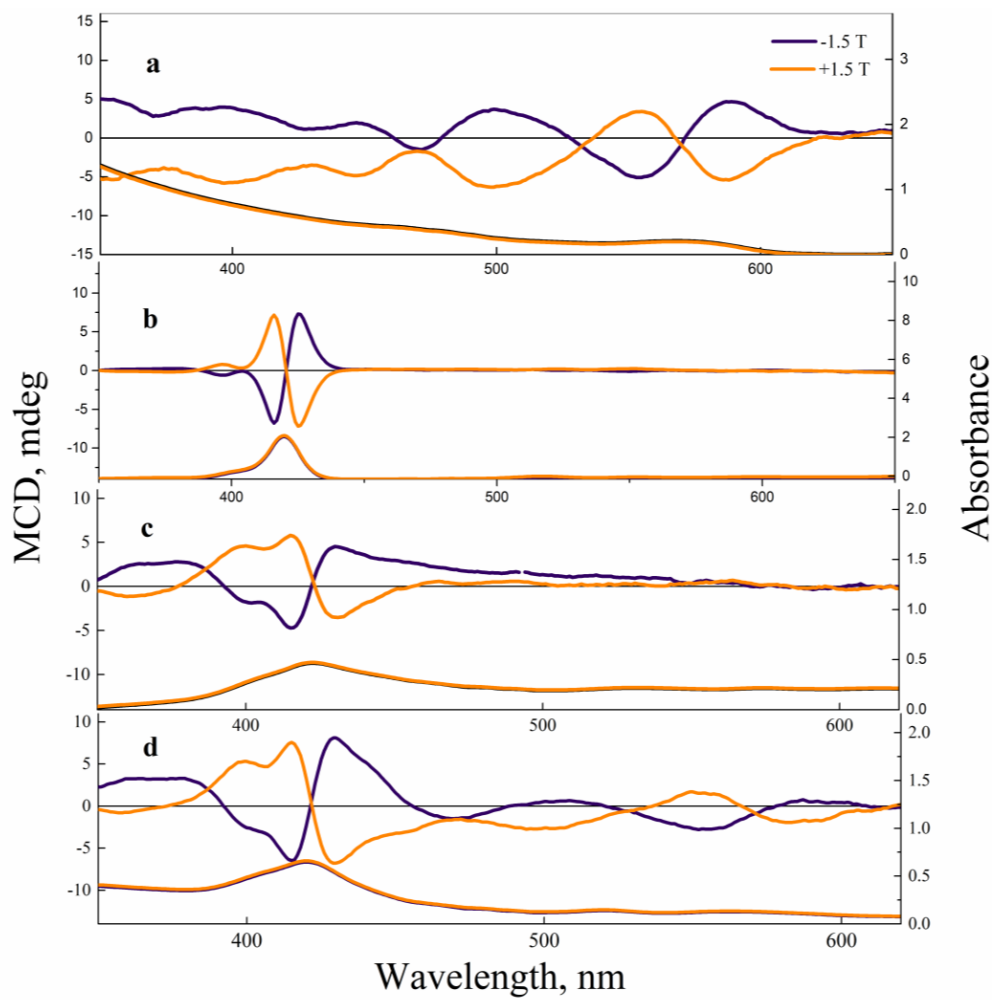

Figure 4. MCD (the top panel) and absorption (bottom panel) spectra of CdSe/ZnS QDs in CS (a) free TPP in CCl$_4$ (b), free TPP in CS (c), and CdSe/ZnS QDs-TPP nanocomposite with the molar ratio $n = 4$ in CS (d).

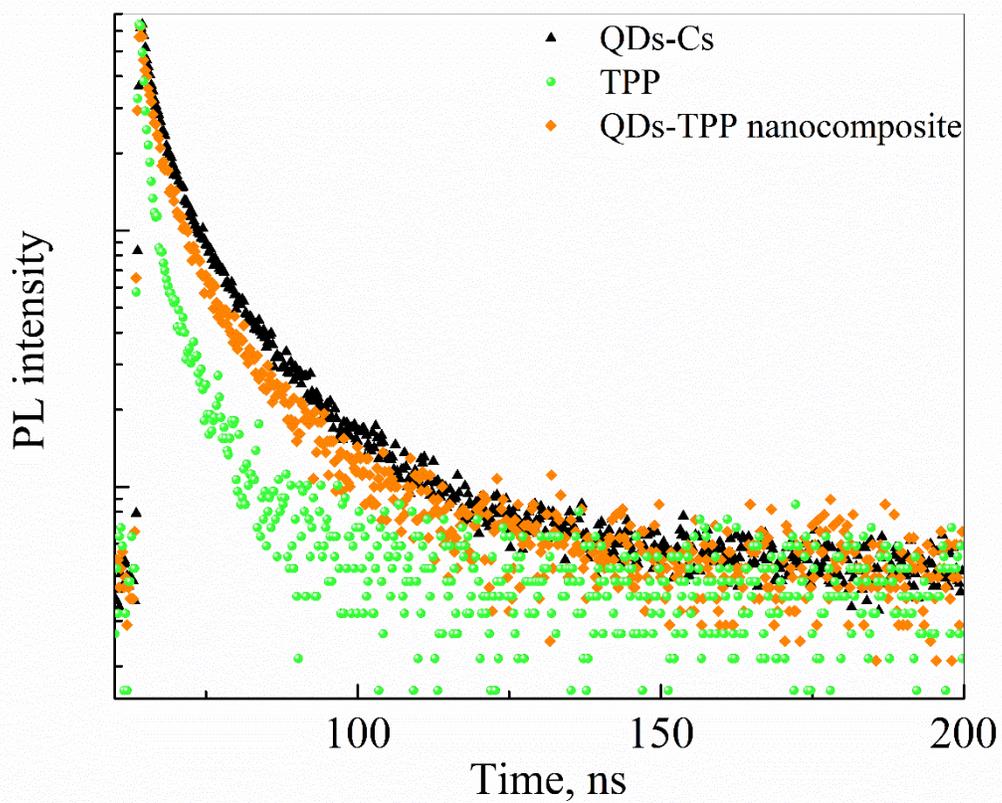

Figure 5. PL decays curves of CdSe/ZnS QDs, TPP, and QDs-TPP nanocomposite dispersed in CS solution, the interference filters were used to split QD and TPP PL signals.

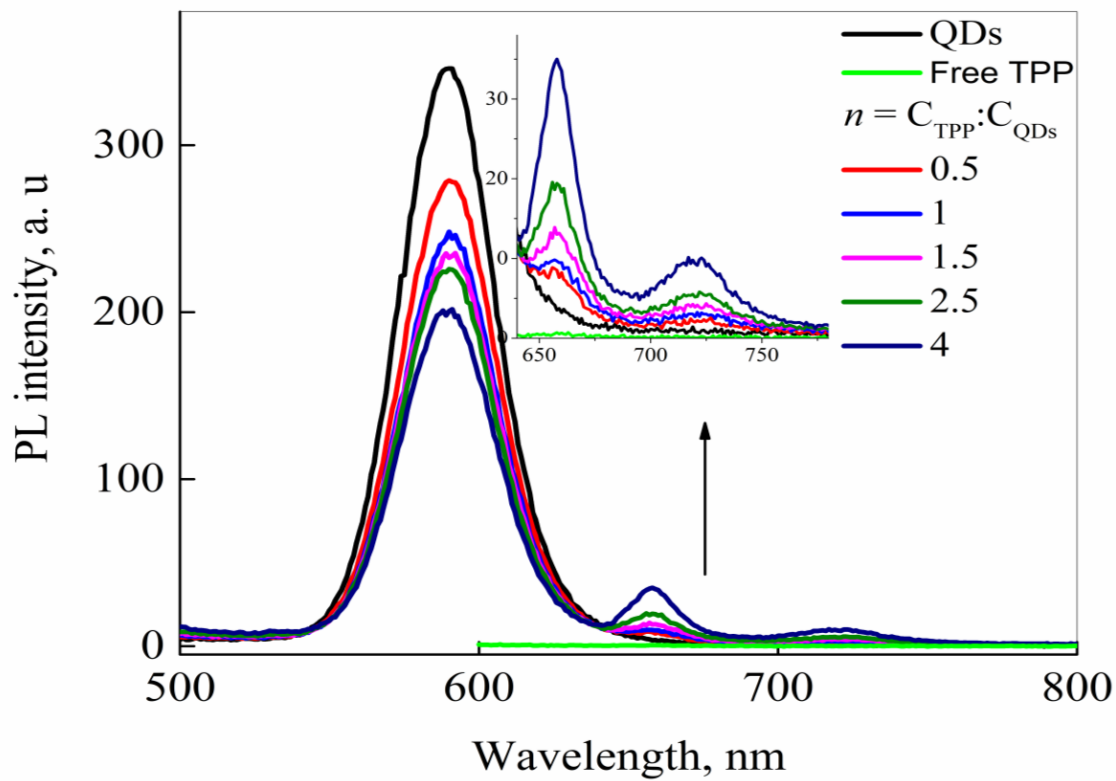

Figure 6. PL spectra of free QDs, free TPP (1.7·10$^{-7}$ M) and QDs-TPP nanocomposites with different *n*; PL excitation wavelength is 460 nm.

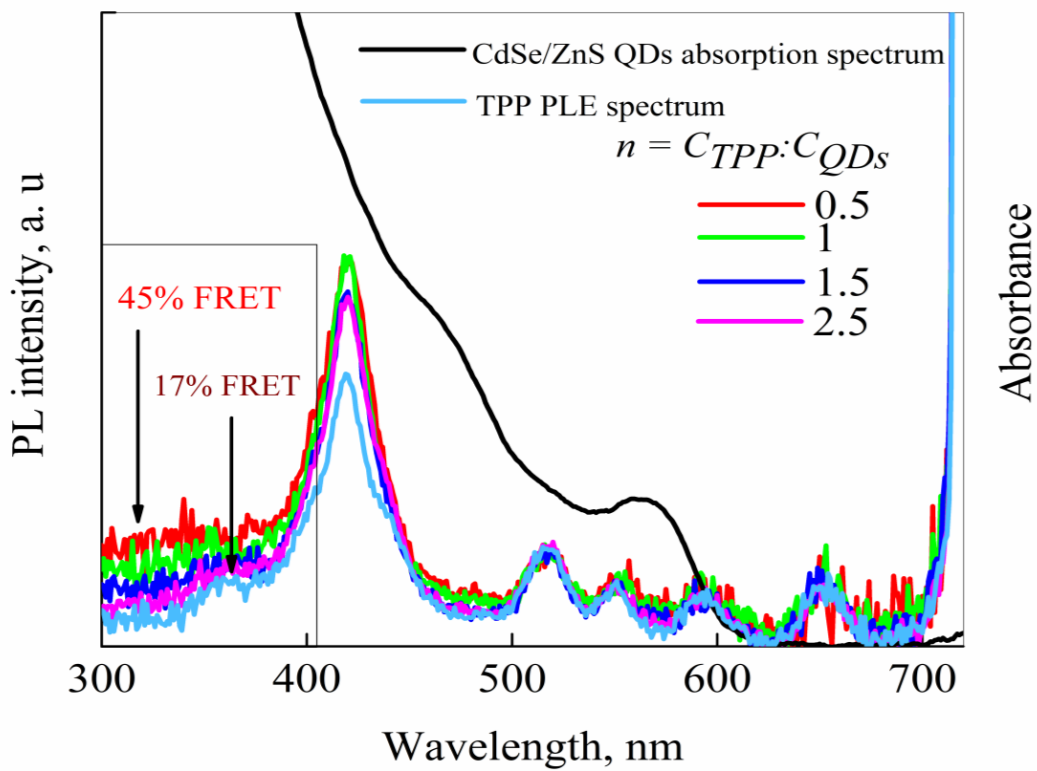

Figure 7. PLE spectra of CdSe/ZnS QDs-TPP nanocomposite with different $n$. PLE spectrum of TPP and QDs absorption spectrum. PL registration wavelength is 725 nm.

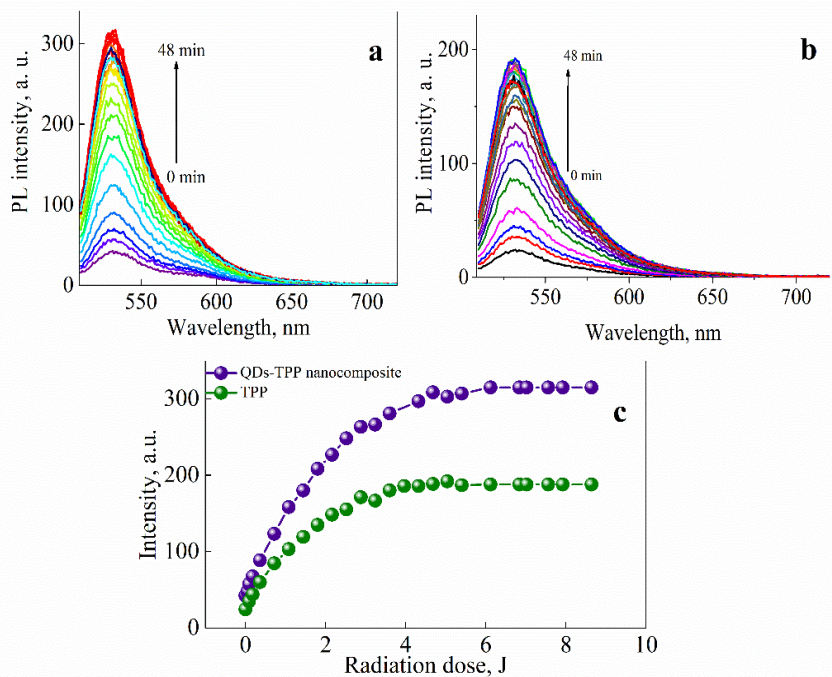

Figure 8. PL spectra of SOSG in (a) QDs-TPP nanocomposite and (b) free TPP in CS excited at 500 nm, with increasing radiation dose (time of irradiation) by visible light at 460 nm; (c) the intensity of SOSG PL emission at 530 nm as a function of radiation dose.